\documentclass[lettersize,journal]{IEEEtran}
\usepackage{amsmath,amsfonts}
\usepackage{algorithmic}
\usepackage{algorithm}
\usepackage{array}
\usepackage[caption=false,font=normalsize,labelfont=sf,textfont=sf]{subfig}
\usepackage{textcomp}
\usepackage{stfloats}
\usepackage{subfig}
\usepackage{balance}
\usepackage{url}
\usepackage{verbatim}
\usepackage{graphicx}
\usepackage{cite}
\usepackage{algorithm}
\usepackage{algorithmic}
\usepackage[colorlinks]{hyperref}
\hypersetup{citecolor=blue}
\usepackage[font=small,skip=2pt]{caption}

\usepackage{enumitem}
\setlist{nosep}

\begin{document}
\IEEEaftertitletext{\vspace{-10mm}}
\title{Peak Sidelobe Suppression in Planar Fluid Antenna Array}

\author{Haoyu Liang, Zhentian Zhang, Yuanhui Wu, Jingyuan Xu, Hao Jiang, Zaichen Zhang
	\thanks{ }
	\thanks{Haoyu Liang, Zhentian Zhang, Zaichen Zhang are with the National Mobile Communications Research Laboratory, Frontiers Science Center for Mobile Information Communication and Security, Southeast University, Nanjing, 210096, China. Zaichen Zhang are also with the Purple Mountain Laboratories, Nanjing 211111, China. (e-mails: zhentianzhangzzt@gmail.com, \{lianghaoyu, zczhang\}@seu.edu.cn). {\em Corresponding Author: Zaichen Zhang.}
	
	
	Hao Jiang is with the School of Cyber Science and Engineering, Southeast University, Nanjing 210096, P. R. China (e-mail: jiang.hao@seu.edu.cn).}
	\thanks{Y. Wu is with the College of Artificial Intelligence, Nanjing University of Information Science and Technology, 210044, China. (e-mails: 202412621447@nuist.edu.cn)}
	\thanks{J. Xu is with the Southeast Unversity, NanJing, Jiangsu, 211189, China. (Email: jingyuanxu@seu.edu.cn)}
		}

\maketitle

\begin{abstract}
	Fluid antenna systems (FAS) have emerged as a promising technology for next-generation
	wireless communications, offering inherent reconfigurability and spatial adaptability.
	A distinctive and practically consequential property of fluid antenna arrays (FAAs) is
	their geometric diversity: by dynamically activating different subsets of spatially
	distributed ports across a dense discrete grid, a FAA can reconfigure its effective
	aperture geometry on demand, thereby unlocking unprecedented spatial degrees of freedom
	for radiation pattern synthesis. Exploiting such geometric flexibility, this paper
	investigates peak sidelobe level (PSLL) minimization in sparse planar FAAs through
	enhanced heuristic optimization. Specifically, an improved genetic algorithm (IGA) is
	proposed to determine the optimal port activation pattern that minimizes the PSLL under
	strict sparsity constraints. The proposed IGA incorporates tournament selection, adaptive
	operator probabilities, a hybrid crossover scheme, multi-point mutation, and an
	elite-pool preservation strategy to improve both convergence speed and solution quality.
	Simulation results demonstrate that the IGA significantly outperforms the canonical GA
	(CGA) in convergence behavior and final PSLL performance, achieving a $4.45\,\mathrm{dB}$
	reduction in sidelobe levels while maintaining a comparable mainlobe width.
\end{abstract}


\section{Introduction}

Fluid antenna systems (FAS) \cite{fas-twc-21, kit_electronic} have catalyzed a paradigm
shift in next-generation wireless networks, establishing foundational theories across
multiple access \cite{MA1,MA2,MA3}, random access \cite{RA1,RA2,RA3}, multi-carrier
systems \cite{OFDM1,OFDM2}, low-latency communications \cite{ll1,ll2}, beamforming
\cite{beam1,beam2}, and novel signal processing \cite{signal}. Early FAS research
primarily adopted a fading-domain perspective, exploiting the unique plateau and
deep-fading characteristics of the channel envelope to extract diversity gains and
improve communication reliability. However, a complementary and arguably more
fundamental dimension of FAS has recently come to the fore: the \emph{geometric
	diversity} inherent in a fluid antenna array (FAA) \cite{beam2, planar_FAA}, which, different to the fading benefits, does not take much overhead to be exploited.

Unlike conventional fixed arrays whose aperture geometry is permanently determined at
fabrication, a FAA allows active ports to dynamically shift across a {\em densely} packed
discrete spatial grid. This reconfigurable geometry directly governs the array's
effective aperture, spatial sampling structure, and spatial frequency content, and
hence its radiation pattern, enabling performance that is not merely improved but
qualitatively distinct from what fixed apertures permit. As established in
\cite{beam2, planar_FAA} for both linear and planar FAA topologies, such geometric
flexibility serves as the enabling mechanism for fine-grained radiation pattern control,
including beamforming gain, directivity shaping, and, most critically for
interference-limited systems, sidelobe suppression \cite{zzt-7, array design}. Among these metrics, peak sidelobe level (PSLL) minimization is of particular
importance, as excessive sidelobes directly degrade adjacent-channel interference
rejection and the overall spatial selectivity of the system. A broad range of heuristic
optimization algorithms has been explored for PSLL reduction, including grey wolf
optimization (GWO) \cite{GWO}, cuckoo search \cite{MCS}, and chicken swarm
optimization \cite{cuckoo search--chicken swarm optimisation algorithm}. 

Although these methods have demonstrated notable improvements in optimization efficiency for
conventional fixed arrays, their direct applicability to the discrete port-selection
nature of an FAA remains severely limited. The dense geometric layout of a planar FAA
introduces a massive combinatorial search space, causing traditional heuristics to
suffer from sluggish convergence and premature entrapment in local optima. Furthermore,
most existing approaches do not explicitly account for the geometry-dependent radiation
behavior of FASs, leaving an important methodological gap. Developing robust
optimization algorithms tailored specifically to navigate the complex geometric
constraints of FAS-based PSLL minimization therefore represents a critical and open
research direction.
\begin{figure}[t!]
	\centering
	\includegraphics[width=0.8\columnwidth]{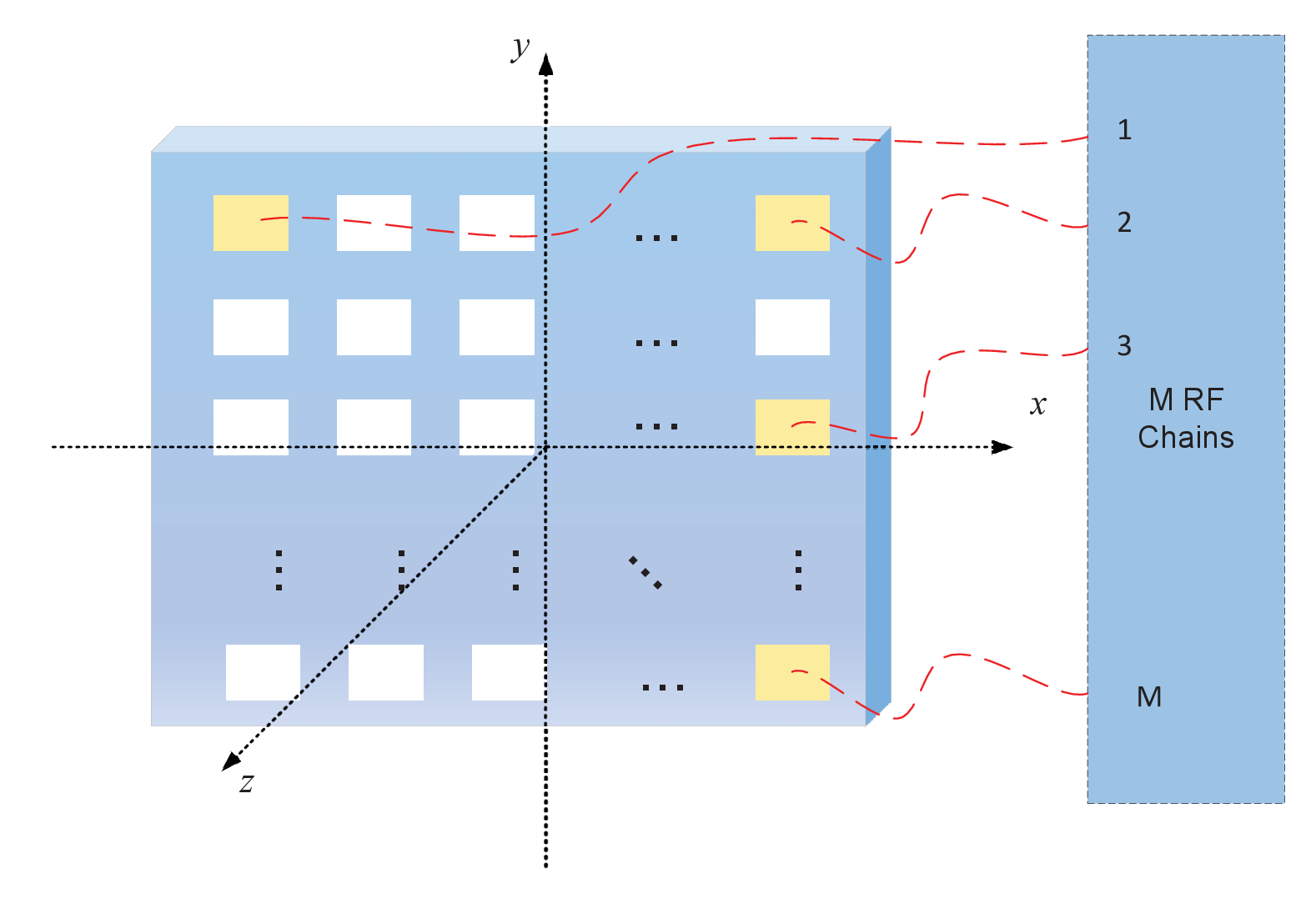}
	\caption{Illustration of FA port array.}
	\label{fig.1}
\end{figure}
The main contributions of this paper are summarized as follows:
\begin{itemize}
	\item The PSLL minimization problem for sparse planar FAAs is formulated as a
	constrained combinatorial optimization task. The objective is to minimize the PSLL
	subject to a cardinality constraint on the number of active ports and a mandatory
	corner-port deactivation constraint, both of which arise naturally from the geometric
	structure of the FAA.
	
	\item A binary-coded canonical genetic algorithm (CGA) is introduced as a baseline
	solver. The CGA employs standard genetic operators (selection, crossover, and
	mutation) together with a constraint repair mechanism to ensure the geometric
	feasibility of all generated solutions.
	
	\item An improved GA (IGA) is proposed to overcome the premature convergence and
	parameter sensitivity of the CGA. The IGA integrates tournament selection, adaptive
	operator probabilities, a hybrid crossover scheme, multi-point mutation, and an
	elite-pool preservation strategy to enhance convergence dynamics and solution
	diversity within the massive FAA search space.
	
	\item Comprehensive simulations are conducted to compare the IGA and CGA. Results
	demonstrate that the IGA achieves markedly faster convergence even for
	high-dimensional port layouts, reduces active port consumption by $88\%$ relative
	to the full array, and improves the PSLL by $4.45\,\mathrm{dB}$ at the cost of only
	a $1.43\,\mathrm{dB}$ reduction in directivity.
\end{itemize}

\section{Configurations and Problem Formulation}
\subsection{Planar FAA Configurations}
Considering a planar FAA, as illustrated in Fig.~\ref{fig.1}, the ports of the planar FAA are arranged on an $N=N_x \times N_y$ grid, where $N_x$ and $N_y$ denote the number of ports along the horizontal and vertical axes, respectively. The inter-port placement gap in the 2-D plane are given by $d_x$ and $d_y$, and the carrier wavelength is denoted as $\lambda$. With limited radio frequency (RF) chains constraint, the planar FAA activates a subset of ports from the total $N$ available ports (i.e., $M\le N$). {\em In general, one would favor the number of ports to be activated as low as possible for the PSLL optimization (i.e., M can be deemed as a systematic constraint or optimization goal.)}

Define $A F(\theta, \varphi)$ as the summation of steering elements, termed as {\em array factor}, containing the on-off patterns of ports and corresponding phase information. The array factor is expressed as
\begin{equation}
AF(\theta, \varphi) 
= \sum_{m=0}^{N_x-1}\sum_{n=0}^{N_y-1} w_{mn}
\exp\Big\{ jk \big[ d_m\Delta_1 + d_n\Delta_2 \big] \Big\} 
\end{equation}
where constants $\Delta_1, \Delta_2$ denotes the angle-oriented phase information (beamforming direction)
\begin{equation}
\begin{aligned}
\Delta_1 &= \sin\theta\cos\varphi-\sin\theta_0\cos\varphi_0, \\
\Delta_2 &= \sin\theta\sin\varphi-\sin\theta_0\sin\varphi_0,
\end{aligned}
\end{equation}
and $w_{mn}\in\{0,1\}$ serves as a binary state coefficient of the $(m,n)$-th port, with $w_{mn}=1$ denoting {\em active} states, otherwise, {\em inactive} states, wave number is dented by $k=2\pi/\lambda$, and $(\theta_0,\varphi_0)$ denotes the desired steering direction, where $\theta_0$ and $\varphi_0$ represent the elevation and azimuth angles, respectively. In this work, the receiver harness the geometrical flexibility by the on-off coefficient optimization where only $M$ ports are to be activated with their binary coefficients $w_{mn}$ equal to one.

\subsection{Optimization Problem Formulation}
Sidelobes represent the unwanted secondary lobes in antenna radiation patterns, where the PSLL serves as a critical metric for evaluating antenna performance. Lower PSLL indicates better power concentration at the main lobe. With planar FAA, we formulate the problem of PSLL minimization into sparse planar array optimization. Define the PSLL expression
\begin{equation}
\Psi \stackrel{\triangle}{=} 20 \log _{10}\left(\frac{\max _{(\theta, \varphi) \in S}|A F(\theta, \varphi)|}{\max |A F(\theta, \varphi)|}\right) 
\end{equation}
where $S$ denotes the sidelobe region, the numerator and the denominator respectively denote the received power of main lobe and the side lobe. Accordingly, the optimization problem can be mathematically formulated as follows:
\begin{subequations}\label{eq.problem}
	\begin{align}
&\min _{w_{m n}} \Psi, \\
& \text { s.t. } \sum_{m=0}^{N_x-1} \sum_{n=0}^{N_y-1} w_{m n} \leq M, w_{m n} \in\{0,1\}
\end{align}
\end{subequations}
where $w_{mn}\in\{0,1\}$
 is the binary variable describing the activation state of the $(m, n)$-th port, and $M$ denotes the maximum permissible count of active ports. {\em The goal is to minimize the PSLL of the antenna array with hardware constraint}. Particularly, $M$ only cells the number of activated ports.

\section{Proposed PSLL Minimization Algorithms}

\begin{figure}[t!]
	\centering
	\includegraphics[width=0.7\columnwidth]{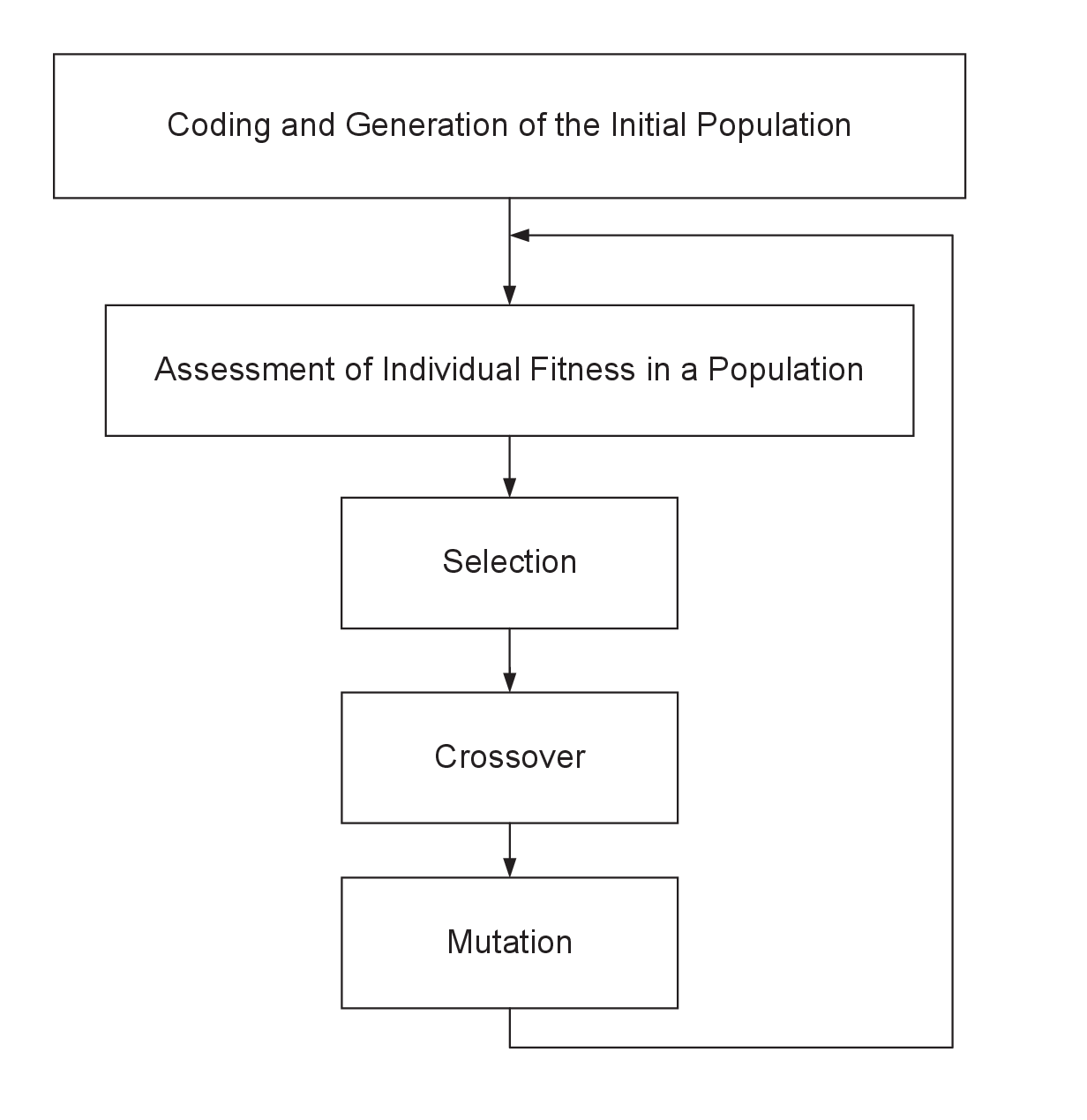}
	\caption{Flowchart of the proposed algorithms.}
	\label{CGA_flow}
\end{figure}

In this section, we present two binary-coded genetic algorithm (GA) variants to optimize 
the port activation pattern of the considered planar FAA subject to practical hardware 
constraints. The first, referred to as the canonical GA (CGA), serves as the performance 
baseline. The second, referred to as the improved GA (IGA), incorporates tournament 
selection, adaptive operator probabilities, a hybrid crossover scheme, multi-point 
mutation, and an elite-pool preservation strategy to overcome the limitations of CGA. 
Both algorithms evolve the port activation configuration through iterative selection, 
crossover, mutation, and constraint repair, with parallel fitness evaluation to accelerate 
computation. The flowchart is illustrated in Fig.~\ref{CGA_flow}.

\subsection{Canonical GA for PSLL Minimization}

The CGA performs population-based evolution relying on three fundamental operators, 
namely selection, crossover, and mutation, and features a concise and explainable 
implementation. The parametric tuple of CGA is defined as
\begin{equation}
	\mathrm{CGA} = \bigl(C,\; E,\; \mathbf{P}_0,\; N_{\mathrm{pop}},\;
	\mathcal{S},\; \mathcal{C},\; \mathcal{M},\; T_{\max}\bigr),
\end{equation}
where $C$ is the binary encoding scheme, $E$ is the fitness evaluation function, 
$\mathbf{P}_0$ is the initial population of size $N_{\mathrm{pop}}$, $\mathcal{S}$ is 
the selection operator, $\mathcal{C}$ is the crossover operator, $\mathcal{M}$ is the 
mutation operator, and $T_{\max}$ is the maximum number of iterations.

\paragraph{Encoding and Initialization}
Binary encoding is adopted to represent the port activation states of the planar FAA. 
By flattening the two-dimensional port matrix row by row, the $i$-th chromosome is 
defined as a binary vector of length $N = N_x \times N_y$,
\begin{equation}
	\mathbf{w}_i = \bigl[w_{i,1},\; w_{i,2},\; \ldots,\; w_{i,N}\bigr],
	\quad w_{i,j} \in \{0,1\},
\end{equation}
where $w_{i,j} = 1$ denotes an activated port. The hardware constraint requires exactly 
$M$ active ports per chromosome, imposing the $\ell_0$-norm constraint 
$\|\mathbf{w}_i\|_0 = M$. Additionally, ports at the four corner indices 
$\mathcal{I}_c = \{1,\; N_y,\; (N_x-1)N_y+1,\; N_x N_y\}$ are prohibited from 
activation throughout the entire evolutionary process, so the feasible set is
\begin{equation}
	\mathcal{W}_c = \Bigl\{\mathbf{w} \in \{0,1\}^N 
	\;\Big|\; \|\mathbf{w}\|_0 = M,\; w_j = 0\; \forall\, j \in \mathcal{I}_c \Bigr\}.
\end{equation}
The initial population $\mathbf{P}_0 = \{\mathbf{w}_1, \ldots, \mathbf{w}_{N_{\mathrm{pop}}}\}$ 
is generated by randomly assigning exactly $M$ ones in each chromosome subject to 
$\mathcal{W}_c$.

\paragraph{Array Factor and Fitness Function}
Given a chromosome $\mathbf{w}$, the physical coordinates of the $j$-th port are
\begin{equation}
	x_j = \Bigl(\lceil j/N_y \rceil - \tfrac{N_x+1}{2}\Bigr)d_x,
	\quad
	y_j = \Bigl((j \bmod N_y) - \tfrac{N_y+1}{2}\Bigr)d_y,
\end{equation}
so that the aperture is centered at the origin. The array factor in the 
$(u,v) = (\sin\theta\cos\varphi,\, \sin\theta\sin\varphi)$ domain is
\begin{equation}\label{eq:AF}
	AF(u,v;\mathbf{w}) = \sum_{j=1}^{N} w_j\,
	\exp\!\Bigl(j2\pi/\lambda\bigl(x_j u + y_j v\bigr)\Bigr),
\end{equation}
evaluated on the visible region $\mathcal{V} = \{(u,v) : u^2+v^2 \le 1\}$. 
The normalized pattern in decibels is
\begin{equation}\label{eq:AFdB}
	\overline{AF}_{\mathrm{dB}}(u,v;\mathbf{w}) = 20\log_{10}
	\frac{|AF(u,v;\mathbf{w})|}{\displaystyle\max_{(u',v')\in\mathcal{V}}
		|AF(u',v';\mathbf{w})|}.
\end{equation}
To evaluate the PSLL without contamination from the main beam, a mainlobe exclusion
region is introduced and the sidelobe search region is defined as
\begin{equation}\label{eq:SLregion}
	\Omega_{\mathrm{SL}} = \bigl\{(u,v)\in\mathcal{V} :|u| \ge \tfrac{2}{N_x}\;\text{or}\;|v| \ge \tfrac{2}{N_x}\bigr\},
\end{equation}
where the threshold $\tfrac{2}{N_x}$ corresponds to the first-null beamwidth of the fully
populated reference array,
ensuring that the mainlobe is entirely excluded from the sidelobe penalty region.
The same definition is applied to all compared methods to guarantee a fair and
consistent PSLL evaluation.

and the PSLL is accordingly evaluated as
\begin{equation}\label{eq:PSLL}
	\Psi(\mathbf{w}) = \max_{(u,v)\in\Omega_{\mathrm{SL}}}\,
	\overline{AF}_{\mathrm{dB}}(u,v;\mathbf{w}).
\end{equation}
Since the GA maximizes a fitness function whereas the optimization objective 
\eqref{eq.problem} minimizes $\Psi$, the fitness of chromosome $\mathbf{w}_i$ is 
defined as
\begin{equation}\label{eq:fitness_CGA}
	F(\mathbf{w}_i) = -\Psi(\mathbf{w}_i),
\end{equation}
so that maximizing $F$ is equivalent to minimizing the PSLL. Fitness evaluations 
across the entire population are executed in parallel to reduce wall-clock computation 
time.

\paragraph{Selection Operator}
Roulette-wheel selection is adopted in CGA. Because raw fitness values may be negative, 
a non-negative transformation is first applied,
\begin{equation}
	\widetilde{F}(\mathbf{w}_i) = \max\bigl(F(\mathbf{w}_i),\; 0\bigr).
\end{equation}
The selection probability of individual $\mathbf{w}_i$ is then
\begin{equation}\label{eq:roulette}
	p_i = \frac{\widetilde{F}(\mathbf{w}_i)}
	{\displaystyle\sum_{k=1}^{N_{\mathrm{pop}}} \widetilde{F}(\mathbf{w}_k)},
\end{equation}
and $N_{\mathrm{pop}}$ parents are drawn with replacement according to 
$\{p_i\}_{i=1}^{N_{\mathrm{pop}}}$ to form the mating pool.

\paragraph{Crossover Operator}
Two-point segment crossover is applied to consecutive parent pairs 
$(\mathbf{P}_1, \mathbf{P}_2)$ with probability $p_c$. Two distinct loci 
$c_1 < c_2$ are drawn uniformly from $\{1,\ldots,N\}$, and offspring 
$\mathbf{O}_1$ and $\mathbf{O}_2$ are generated by exchanging the gene segment 
within $[c_1, c_2]$:
\begin{equation}\label{eq:crossover_CGA}
	\mathbf{O}_1[c_1\!:\!c_2] = \mathbf{P}_2[c_1\!:\!c_2],\quad
	\mathbf{O}_1[\mathrm{else}] = \mathbf{P}_1[\mathrm{else}],
\end{equation}
and symmetrically for $\mathbf{O}_2$. If no crossover event occurs, offspring are 
identical copies of their respective parents.

\paragraph{Mutation Operator}
Single-bit-flip mutation is applied to each offspring with probability $p_m$. 
A position $r$ is drawn uniformly from $\{1,\ldots,N\}$, and the corresponding 
bit is complemented,
\begin{equation}\label{eq:mutation_CGA}
	o_{i,r} \leftarrow 1 - o_{i,r}.
\end{equation}

\paragraph{Repair Mechanism}
Crossover and mutation may violate the cardinality constraint or activate a corner 
port. Let $K' = \|\mathbf{o}_i\|_0$ denote the active-port count of the unrepaired 
offspring. If $K' > M$, then $K'-M$ randomly chosen active bits are set to zero. 
If $K' < M$, then $M-K'$ randomly chosen inactive bits are set to one. Any active 
corner port in $\mathcal{I}_c$ is subsequently forced to zero, with the count 
shortfall compensated by activating a randomly selected non-corner inactive bit, 
ensuring $\mathbf{o}_i \in \mathcal{W}_c$ after repair.

\paragraph{Elite Preservation and Termination}
After the repair step, the offspring individual with the lowest fitness is replaced 
by the globally best chromosome $\mathbf{w}^*$ found so far, preventing evolutionary 
degradation. If the $\ell$-th individual has
\begin{equation}
	\ell = \arg\min_{1 \le i \le N_{\mathrm{pop}}} F(\mathbf{w}_i),
\end{equation}
the updated population for the next generation is
\begin{equation}
	\mathbf{P}_{t+1} = \bigl(\text{repaired offspring}\bigr)
	\setminus \{\mathbf{o}_\ell\} \cup \{\mathbf{w}^*\}.
\end{equation}

The algorithm terminates when the iteration counter reaches $T_{\max}$, 
and $\mathbf{w}^*$ is returned as the optimal port activation pattern.

\subsection{Improved GA for PSLL Minimization}

To address the limitations of CGA, namely premature convergence and the imbalance 
between global exploration and local exploitation, an IGA is proposed. While retaining 
the binary encoding, fitness definition \eqref{eq:fitness_CGA}, and repair mechanism, 
IGA replaces or enhances four key components: tournament selection, adaptive operator 
probabilities, a hybrid crossover scheme, and an elite-pool preservation strategy.

\paragraph{Tournament Selection}
Roulette-wheel selection is susceptible to premature convergence when fitness values 
are closely clustered. IGA instead adopts tournament selection: at each selection step, 
$k$ individuals are sampled uniformly at random from the current population to form a 
tournament set $\mathcal{T}$, and the individual with the highest fitness is selected 
as a parent,
\begin{equation}\label{eq:tournament}
	\mathbf{w}_{\mathrm{parent}} =
	\arg\max_{\mathbf{w}_c \in \mathcal{T}} F(\mathbf{w}_c).
\end{equation}
This procedure is repeated $N_{\mathrm{pop}}$ times to fill the mating pool. 
Tournament selection exerts selection pressure proportional to rank rather than 
absolute fitness value, thereby reducing the risk of premature convergence while 
maintaining population diversity, and it requires no fitness truncation.

\paragraph{Adaptive Operator Probabilities}
Fixed operator probabilities cannot dynamically balance exploration and exploitation 
across evolutionary stages. IGA implements linearly decreasing schedules for both 
crossover and mutation probabilities,
\begin{subequations}
	\begin{align}
	p_c(t) &= p_{c,\max} - \bigl(p_{c,\max} - p_{c,\min}\bigr)\frac{t}{T_{\max}},\label{eq:adaptive_pc}\\
		p_m(t) &= p_{m,\max} - \bigl(p_{m,\max} - p_{m,\min}\bigr)\frac{t}{T_{\max}},\label{eq:adaptive_pm}
	\end{align}
\end{subequations}
where $t$ is the current iteration index. The default values in the implementation 
are $p_{c,\max} = 0.9$, $p_{c,\min} = 0.6$, $p_{m,\max} = 0.1$, and 
$p_{m,\min} = 0.01$. This dual-adaptive schedule promotes broad exploration of the 
solution space in early stages and shifts toward fine-grained local exploitation as 
the population matures.

\paragraph{Hybrid Crossover Scheme}
For each parent pair $(\mathbf{P}_1, \mathbf{P}_2)$, crossover is activated with 
probability $p_c(t)$. Conditioned on crossover being performed, two-point segment 
crossover \eqref{eq:crossover_CGA} is executed with probability $0.5$ to preserve 
contiguous local genetic structures. Otherwise, uniform crossover is executed with the 
remaining probability to enhance global search capability. In uniform crossover, a 
binary mask vector $\mathbf{m} \in \{0,1\}^N$ is generated with each entry drawn 
independently from $\mathrm{Bernoulli}(0.5)$. The offspring are produced by the 
element-wise operations
\begin{equation}\label{eq:uniform_crossover}
	\begin{aligned}
			&\mathbf{O}_1 = \mathbf{m} \odot \mathbf{P}_2 + (\mathbf{1} - \mathbf{m}) \odot \mathbf{P}_1,\\
		&\mathbf{O}_2 = \mathbf{m} \odot \mathbf{P}_1 + (\mathbf{1} - \mathbf{m}) \odot \mathbf{P}_2,
	\end{aligned}
\end{equation}
where $\odot$ denotes element-wise multiplication and $\mathbf{1}$ is the all-ones 
vector of length $N$.

\paragraph{Multi-Point Mutation}
To avoid the inefficiency of single-bit perturbation on a chromosome of length $N$, 
IGA applies multi-point bit-flip mutation. When mutation is triggered with probability 
$p_m(t)$, the number of simultaneously flipped bits is set to
\begin{equation}\label{eq:mutate_num}
	L = \max\!\bigl(1,\; \mathrm{round}(0.05\,N)\bigr).
\end{equation}
A set $\mathcal{P}_{\mathrm{mut}}$ of $L$ distinct positions is selected uniformly at 
random, and the corresponding bits are inverted simultaneously,
\begin{equation}\label{eq:mutation_IGA}
	w_{i,j} \leftarrow 1 - w_{i,j}, \quad \forall\, j \in \mathcal{P}_{\mathrm{mut}}.
\end{equation}
This multi-point scheme increases the probability of escaping poor local structures in 
large-dimensional binary search spaces.

\paragraph{Elite-Pool Preservation}
Instead of retaining only a single best individual, IGA maintains an elite pool of 
size $n_e = 5$. The $n_e$ highest-fitness individuals from the current population are 
collected into the elite set
\begin{equation}\label{eq:elite}
	\mathcal{E}(t) = \bigl\{\mathbf{w} \in \mathbf{P}(t) :
	F(\mathbf{w}) \text{ is among the top } n_e \text{ values}\bigr\}.
\end{equation}
Following fitness evaluation of the offspring population, the $n_e$ individuals with 
the lowest fitness values are deterministically overwritten by $\mathcal{E}(t)$ before 
the next iteration begins. This mechanism guarantees the monotone non-decreasing 
improvement of the best fitness over iterations while injecting multiple high-quality 
genetic structures into the new population to guide subsequent search.

\section{Simulation Results}

In this section, simulation results are presented to evaluate the performance of the 
proposed IGA for PSLL minimization in sparse planar FA arrays. The simulation parameters 
are summarized in Table~\ref{tab:sim_params11}. The IGA is compared against the CGA 
with respect to both convergence behavior and final PSLL performance.

\begin{table}[t!]
	\caption{Simulation Parameters}
	\centering
	\renewcommand{\arraystretch}{1.2}
	\begin{tabular}{r|l}
		\hline
		\textbf{Parameter} & \textbf{Value} \\ \hline
		Total number of ports $N$& 2500           \\
		Port spacing $d_x = d_y$                                        & $0.5\lambda$\\
		Carrier frequency $f_c$                                         & 3.5~GHz        \\
		Number of active ports $M$                                      & 300            \\
		Population size $N_{\mathrm{pop}}$                              & 100            \\
		Maximum iterations $T_{\max}$                                   & 300            \\
		CGA crossover probability $p_c$                                 & 0.6            \\
		CGA mutation probability $p_m$                                  & 0.1            \\		
		IGA adaptive crossover bounds $p_{c,\max}$,\,$p_{c,\min}$& 0.9,\,0.6\\
		IGA adaptive mutation bounds $p_{m,\max}$,\,$p_{m,\min}$& 0.1,\,0.01\\\hline
	\end{tabular}
	\label{tab:sim_params11}
\end{table}

\begin{figure}[t!]
	\centering
	\includegraphics[width=0.8\columnwidth]{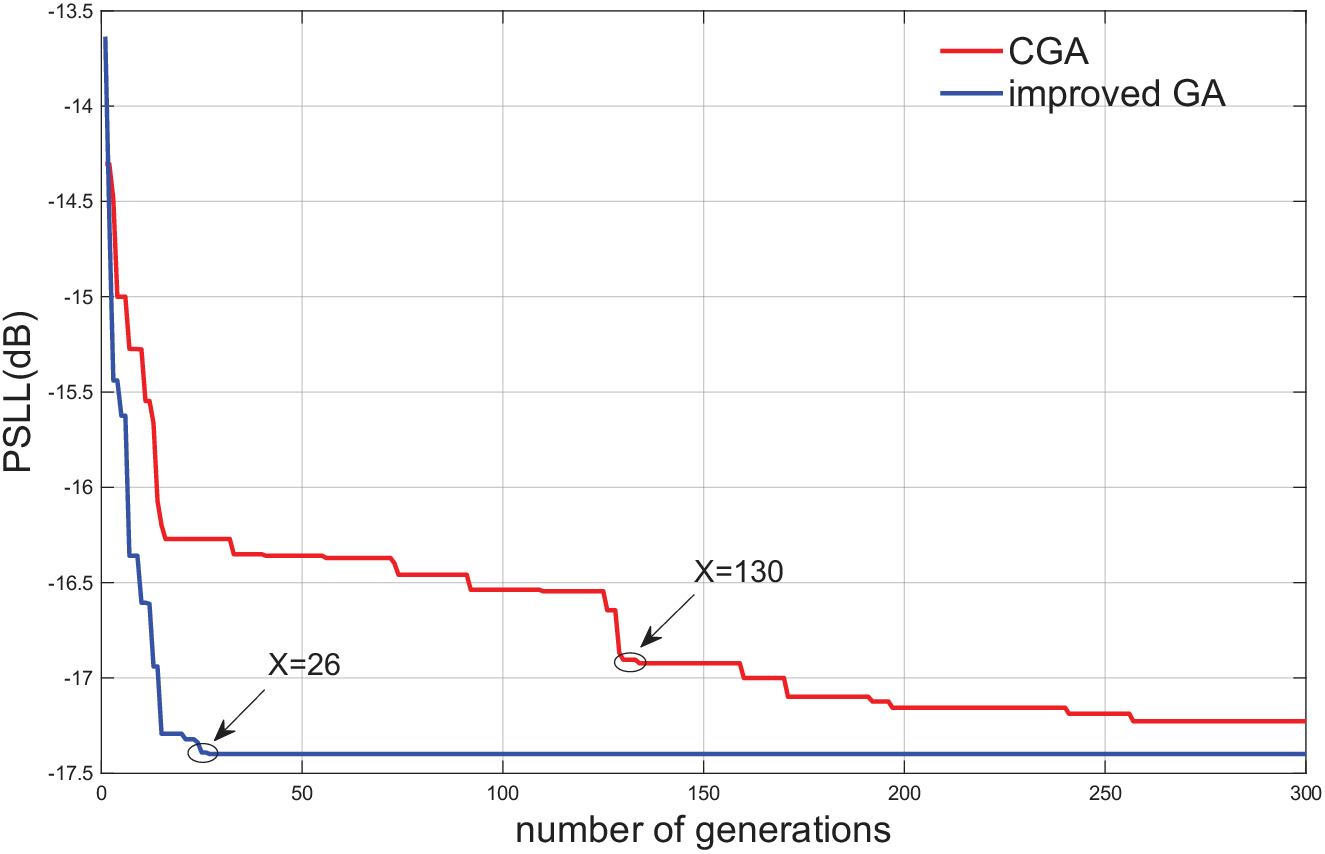}
	\caption{Convergence curves of the proposed IGA and the baseline CGA.}
	\label{fig3}
	\vspace{-4mm}
\end{figure}

As depicted in Fig.~\ref{fig3}, the proposed IGA exhibits a substantially faster 
convergence rate than the CGA. The PSLL decreases rapidly during the early iterations, 
reflecting the IGA's enhanced capacity to explore the high-dimensional binary solution 
space. In contrast, the CGA converges more slowly and shows signs of stagnation, 
suggesting insufficient population diversity maintenance and susceptibility to local 
optima. These observations confirm that the combined enhancements of tournament
selection, adaptive operator probabilities, and elite-pool preservation collectively 
guide the search more effectively toward high-quality array configurations while 
mitigating premature convergence.

\begin{figure}[t!]
	\centering
	\subfloat[Direction map of the full array.]{\includegraphics[width=0.5\columnwidth]{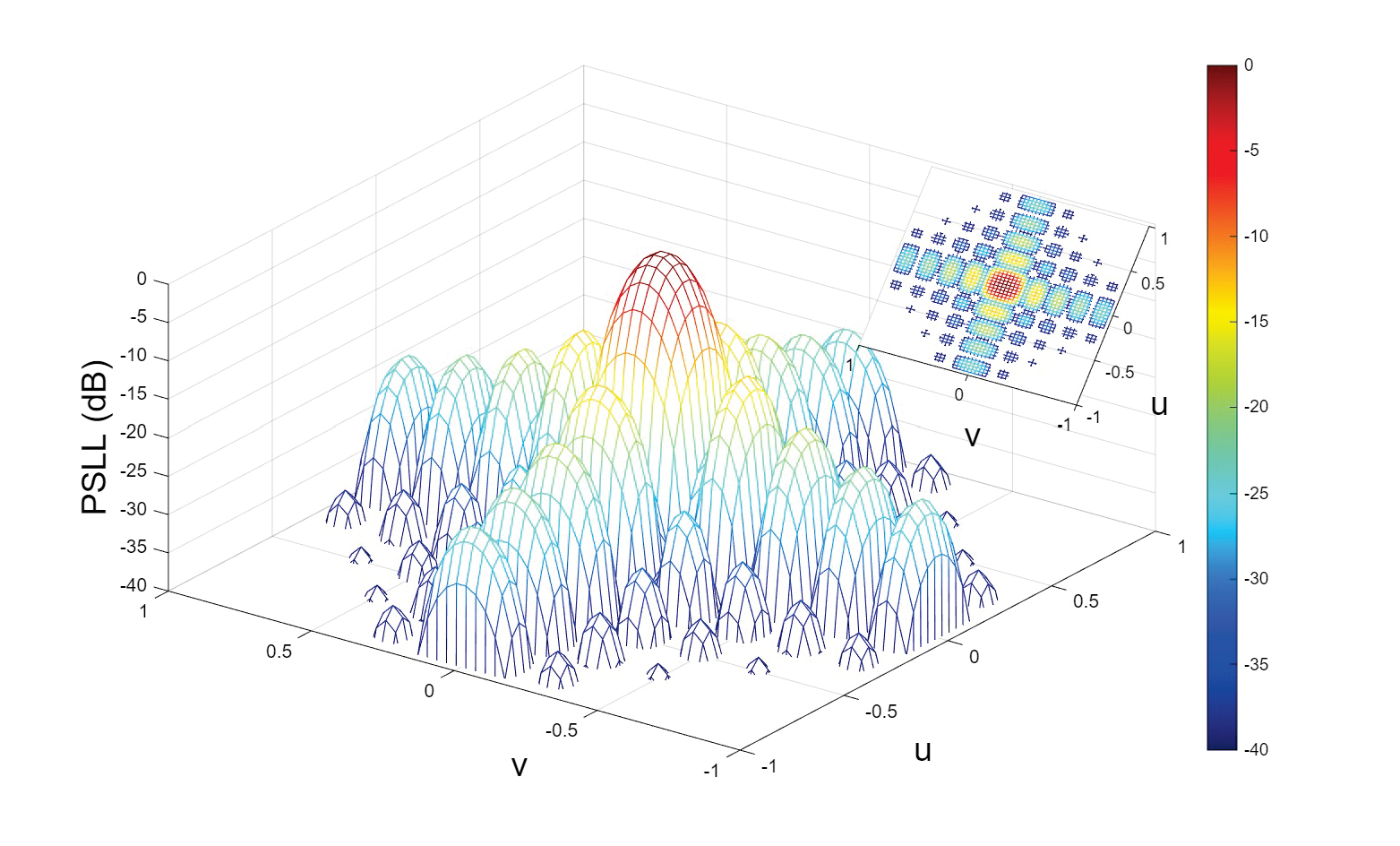}\label{fig4}}
	\hfill
	\subfloat[Direction map of the IGA-optimized sparse array.]
	{\includegraphics[width=0.5\columnwidth]{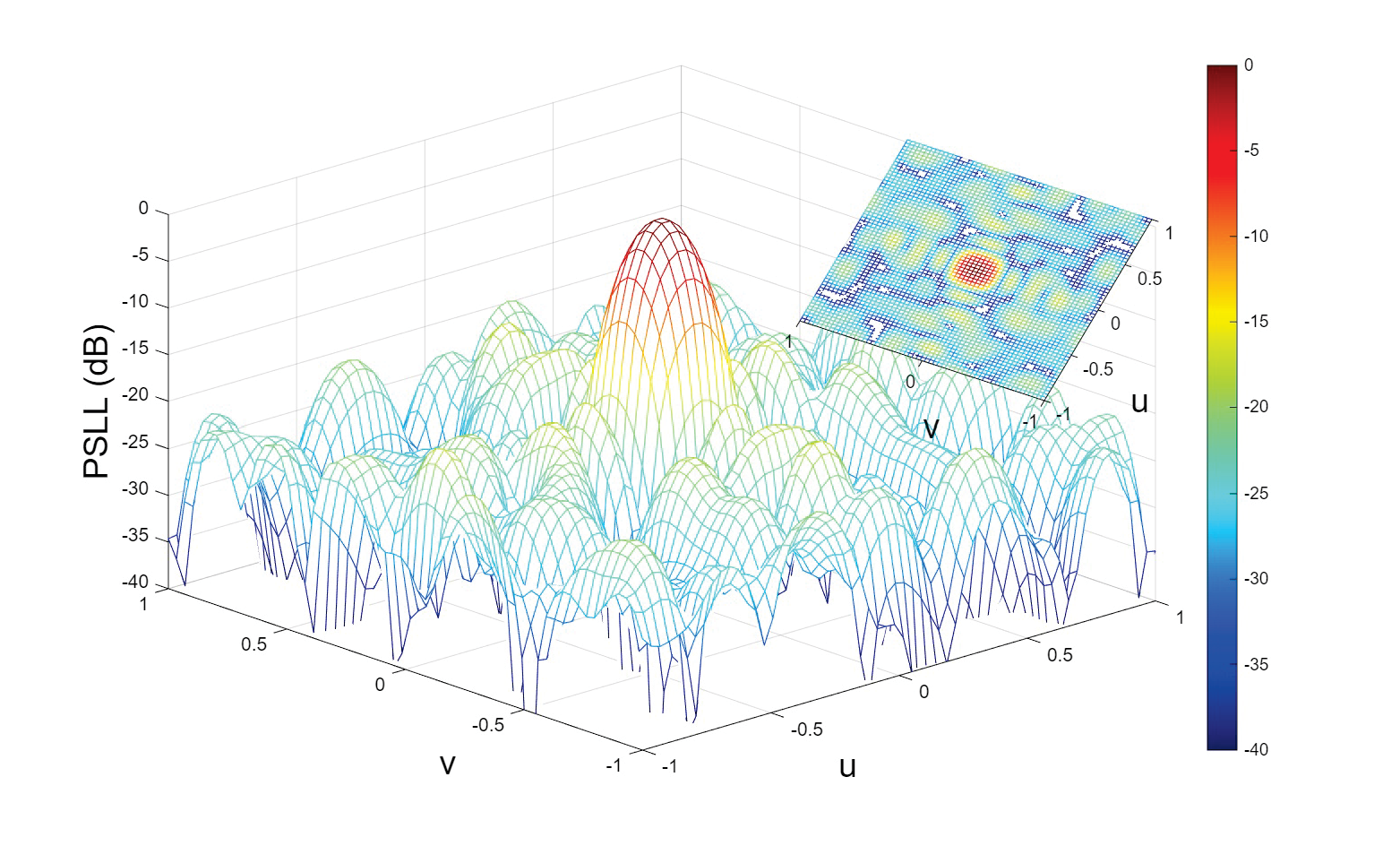}\label{fig5}}
	\caption{Comparison of direction maps between the full array and the 
		IGA-optimized sparse array.}
	\label{fig:direction_maps}
\end{figure}

\begin{table}[t!]
	\centering
	\caption{Performance Comparison Between the Full Array and the IGA-Optimized 
		Sparse Array}
	\label{tab:array_performance}
	\renewcommand{\arraystretch}{1.2}
	\begin{tabular}{ccc}
		\hline
		\textbf{Metric} & \textbf{Full Array} & \textbf{IGA (Proposed)} \\ \hline
		Number of active ports& 2500     & 300      \\
		PSLL (dB)                & $-13.15$ & $-17.56$ \\
		Directivity (dB)         & 24.78    & 23.35    \\
		Port saving rate (\%)    & ---& 88       \\ \hline
	\end{tabular}
	\vspace{-2mm}
\end{table}

The full array, with all 2500 ports activated, yields a PSLL of approximately
$-13.15\,\mathrm{dB}$, with sidelobes exhibiting a symmetric distribution in the 
$(u,v)$ domain, as shown in Fig.~\ref{fig4}. The IGA-optimized sparse array achieves a 
PSLL of approximately $-17.56\,\mathrm{dB}$ (Fig.~\ref{fig5}), corresponding to a 
sidelobe suppression of $4.45\,\mathrm{dB}$. The optimized pattern exhibits a more 
uniform sidelobe distribution with no pronounced high-sidelobe regions, indicating 
effective spatial suppression across the visible region.

\begin{figure}[t!]
	\centering
	\subfloat[Radiation pattern cut along the $u$-axis.]
	{\includegraphics[width=0.5\columnwidth]{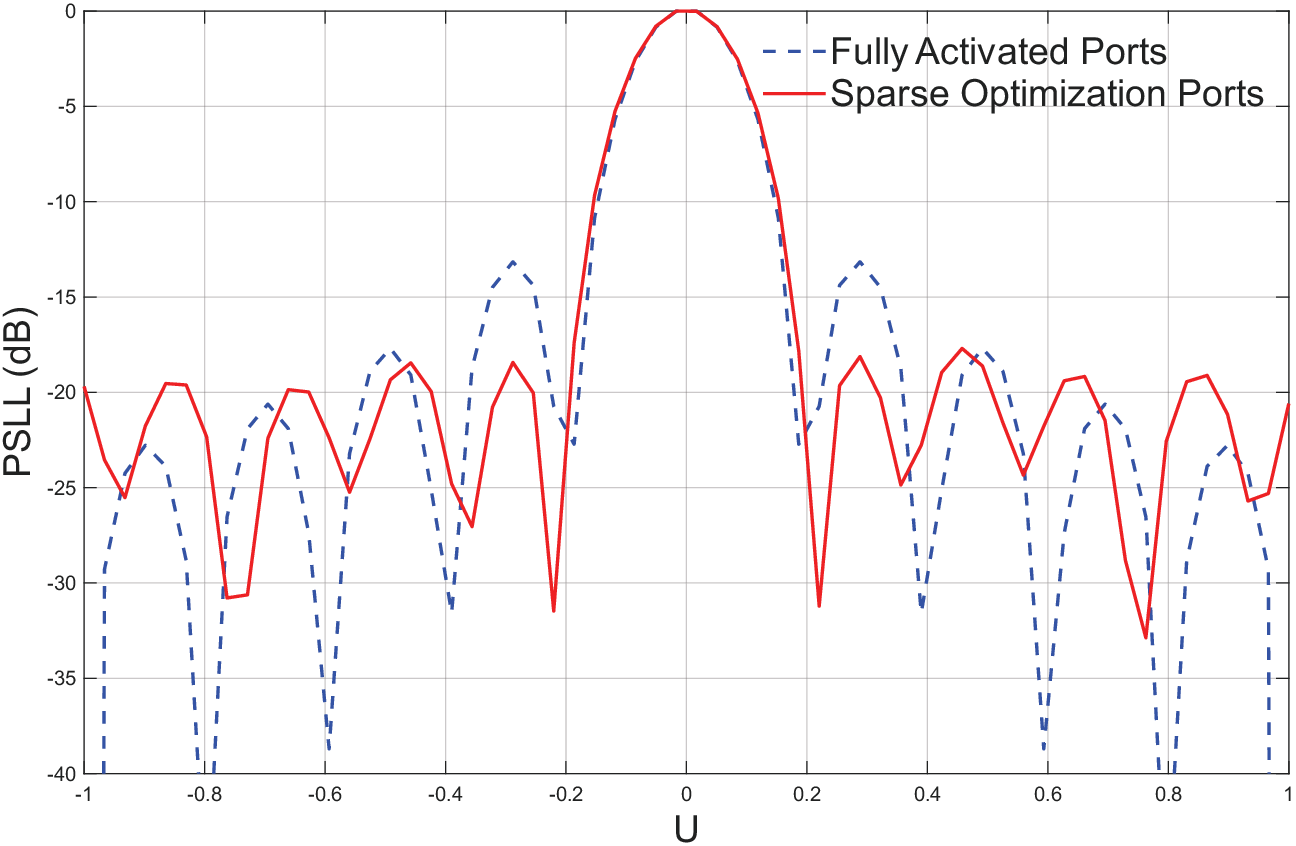}\label{fig6}}
	\hfill
	\subfloat[Radiation pattern cut along the $v$-axis.]
	{\includegraphics[width=0.48\columnwidth]{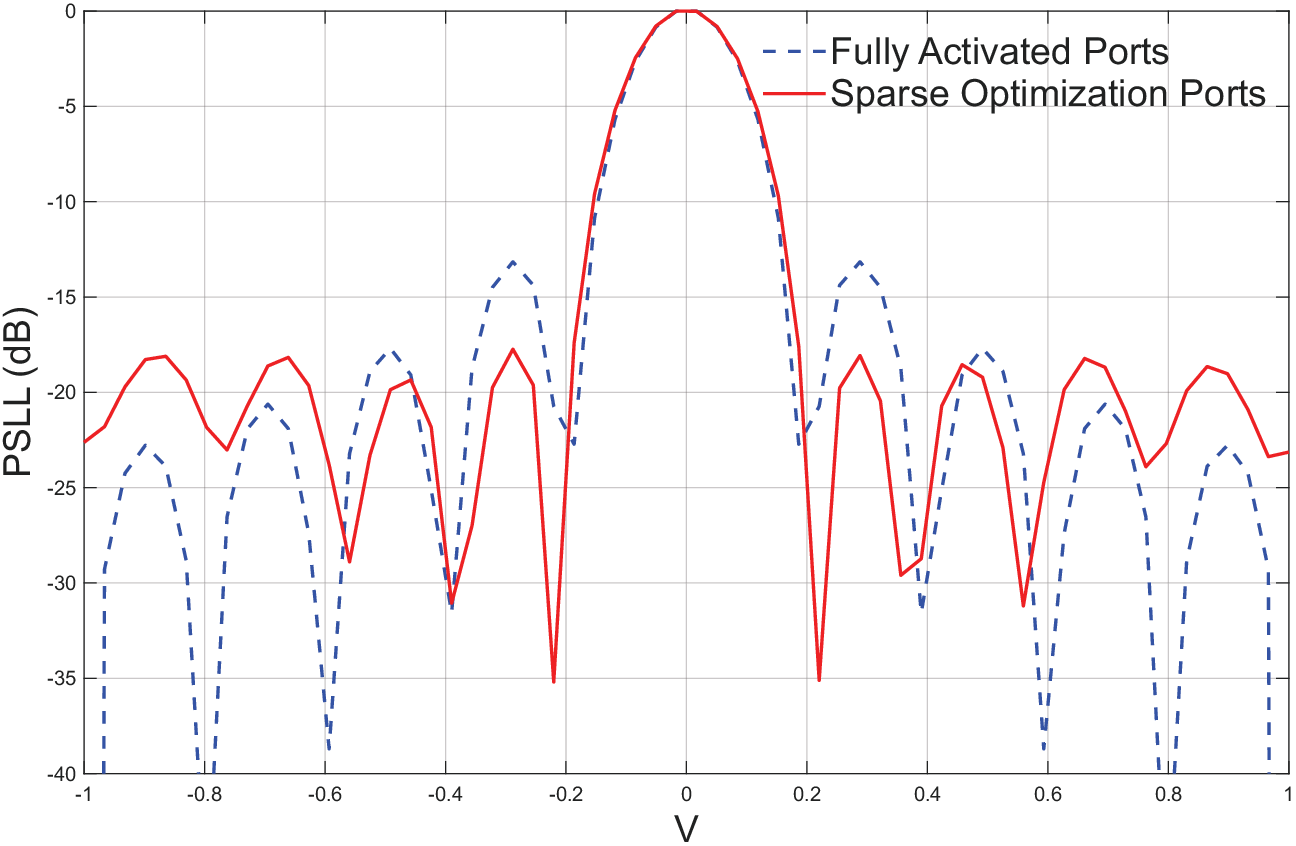}\label{fig7}}
	\caption{Radiation pattern cuts of the IGA-optimized sparse array along the 
		$u$- and $v$-axes.}
	\label{fig:pattern_cuts}
\end{figure}

Fig.~\ref{fig6} and Fig.~\ref{fig7} plot the radiation pattern cuts along the $u$- and 
$v$-axes, respectively. Both orthogonal cross-sections confirm that the IGA-optimized 
sparse array preserves a mainlobe width nearly identical to that of the full array while 
achieving substantially lower sidelobe levels across the entire visible region. This 
demonstrates that effective sidelobe suppression is attained without incurring any 
measurable penalty in mainlobe resolution.

The quantitative results are consolidated in Table~\ref{tab:array_performance}. The IGA 
reduces the number of active ports from 2500 to 300, realizing a port saving rate of 
$88\%$. Despite this $8.3\times$ reduction in active RF chains, the optimized array 
achieves a PSLL of $-17.56\,\mathrm{dB}$, surpassing the full array by
$4.45\,\mathrm{dB}$, while the directivity decreases only marginally from
$24.78\,\mathrm{dB}$ to $23.35\,\mathrm{dB}$ ($1.43\,\mathrm{dB}$ degradation). This 
trade-off confirms that judicious sparse port placement can simultaneously achieve 
significant sidelobe suppression and drastic hardware simplification at the cost of a 
strictly bounded directivity penalty.

\balance
\section{Conclusion}
This paper proposed an IGA for PSLL minimization in sparse planar FAAs, formulated as
a constrained combinatorial optimization problem. Relative to the CGA baseline, the IGA
integrates tournament selection, adaptive operator probabilities, a hybrid crossover
scheme, multi-point mutation, and elite-pool preservation, effectively mitigating
premature convergence and achieving superior optimization performance. The optimized
300-port sparse array reduces RF chain consumption by $88\%$ against the 2500-port full
array, improves the PSLL by $4.45\,\mathrm{dB}$, and incurs only a $1.43\,\mathrm{dB}$
directivity loss. Future work will extend this framework to non-uniform FAA spatial
grids, multi-beam synthesis, and physical prototype validation.

\end{document}